# A THEORETICAL LOOK AT ORDINAL CLASSIFICATION METHODS BASED ON REFERENCE SETS COMPOSED OF CHARACTERISTIC ACTIONS

**Eduardo Fernández**
Universidad Autónoma de Coahuila
eddyf171051@gmail.com

**Jorge Navarro***
Universidad Autónoma de Sinaloa
jnavarro@uas.edu.mx
- **Corresponding author**

**Efraín Solares**
Universidad Autónoma de Coahuila
efrain.solaresl@gmail.com

**Abstract:**

From a theoretical view, this paper addresses the general problem of designing ordinal classification methods based on comparing actions with subset of actions, which are representative of their classes (categories). The basic demand of the proposal consists in setting a relational system ($D, S$), where $S$ is a reflexive relation compatible with the preferential order of the set of classes, and $D$ is a transitive relation such that $D$ is a subset of $S$. Different ordinal classification methods can be derived from diverse model of preferences fulfilling the basic conditions on $S$ and $D$. Two complementary assignment procedures compose each method, which correspond through the transposition operation and should be used complementarily. The methods work under relatively slight conditions on the representative actions and satisfy several fundamental properties. ELECTRE TRI-nC, INTERCLASS-nC, and the hierarchical ELECTRE TRI-nC with interacting criteria, can be considered as particular cases of this general framework.

## 1 Introduction

Classification is one of the most common processes of human mind. It is related to the process of assigning objects (or alternatives, actions) to certain pre-existing classes. We are interested here in ordinal classification problems. Unlike nominal classification, in ordinal classification the decision maker (DM) assigns objects to elements of a set of ordered classes (also called categories in the related literature). When the objects or actions are described by several (or many) evaluation criteria, the DM faces the so-called multi-criteria ordinal classification problem.

Muti-Criteria Decision Aid (MCDA) researchers have developed a plethora of multi-criteria ordinal classification methods. Basically, these differ from two aspects: a) the underlying preference model; and b) the way in which classes are characterized.

From the point of view of the underlying decision model, most of the methods obey to one of three main paradigms:

- The functional paradigm, based on the construction of value functions (*e.g.*, Jacquet-Lagrèze, 1995; Zopounidis and Doumpos, 2000; Köksalan and Ulu, 2003; Köksalan and Özpeynirci, 2009; Greco *et al*., 2010; Bugdaci *et al*., 2013);
- Symbolic methods connected with Artificial Intelligence (*e.g.*, Greco *et al*., 2001; Błaszczyński *et al*., 2007; Fortemps *et al*., 2008; Dembczyński *et al*., 2009); and
- Construction of outranking relations (the relational paradigm) (*e.g.*, Massaglia and Ostanello, 1991; Yu, 1992; Perny, 1998; Belacel, 2000; Tervonen *et al*., 2009; Almeida-Dias *et al.,* 2010; Fernandez and Navarro, 2011; Bouyssou and Marchant, 2015; Meyer and Olteanu, 2019; Bouyssou *et al*., 2020; Fernández *et al*., 2020).

Whatever the model of preferences, classes should be characterized in some way. There are two main approaches:

i) Using limiting actions that describe boundaries between adjacent classes (e.g., Roy and Bouyssou, 1993; Perny, 1998; Araz and Ozkarahan, 2007; Nemery and Lamboray, 2008; Ishizaka *et al*., 2012; Bouyssou and Marchant, 2015; Fernández *et al*., 2017, Bouyssou *et al*., 2020; Fernández *et al*., 2020);

ii) Through decision examples (or reference actions) whose classification is (or may be) known (e.g. Jacquet-Lagrèze, 1995; Zopounidis and Doumpos, 2000; Greco *et al*., 2001; Köksalan and Ulu, 2003; Doumpos and

Zopounidis, 2004; Nemery and Lamboray, 2008; Fernandez et al., 2008, 2009; Köksalan and Özpeynirci, 2009; Köksalan et al., 2009; Almeida-Dias et al., 2010, 2012; Greco et al., 2010; Bugdaci et al., 2013; Izhizaka and Nemery, 2014; Kadzinski et al., 2015; Meyer and Olteanu, 2019; Fernández et al., 2020).

Within the methods in point i), the most outstanding one is ELECTRE TRI (Yu, 1992), later renamed ELECTRE TRI-B (Almeida-Dias et al., 2010). ELECTRE TRI-B defines the boundary between two contiguous classes using a single action. This method was generalized in (Fernández et al., 2017) to create ELECTRE TRI-nB, where several actions can be used to characterize the boundary between each pair of contiguous classes; the idea is that multiple actions provide more informed decisions. Both ELECTRE TRI-B and ELECTRE TRI-nB are based on two different assignment procedures, so-called pseudo-conjunctive and pseudo-disjunctive procedures. Very important structural properties (Unicity, Homogeneity, Independence, Conformity, Monotonicity, and Stability) (see Roy and Bouyssou, 1993) are fulfilled by these procedures. With slight modifications, these properties were adapted and reformulated by Almeida-Dias et al. (2010) in the context of ELECTRE TRI-C. The non-fulfillment of some of these properties could be a serious drawback of other multi-criteria ordinal classification approaches.

Even when the methods based on limiting actions have been widely accepted, they have also received critics regarding the difficulty of defining this type of actions. This is particularly the case when the DM does not have a precise notion about the boundary between adjacent classes. Some authors have shown serious doubts about the existence of these boundaries in real-world situations (e.g., Almeida-Dias et al., 2010; Köksalan et al., 2009). Additionally, ELECTRE TRI-B and its variants have been criticized from a deep theoretical point of view. Roy (2002), Bouyssou and Marchant (2015), and Bouyssou et al. (2020) showed that the pseudo-conjunctive and pseudo-disjunctive procedures do not correspond via the transposition operation. This operation refers to inverting the ordering of the classes and the preference directions of the criteria, such that the conclusions obtained after the inversions should not differ from the conclusions obtained before them (Bouyssou and Marchant, 2015). The value of the information obtained from the statement "$x$ outranks a limiting action $b$" must be equal to that provided by the statement "a limiting action $b$ outranks $x$"; thus, both statements should be considered when assigning $x$ to a class. So, if a multi-criteria ordinal classification procedure has an "image" method through the transposition operation, both methods should be used conjointly. Bouyssou and Marchant (2015), and Bouyssou et al. (2020) have proposed variants of ELECTRE TRI-B and ELECTRE TRI-nB which are symmetric with respect to the transposition operation. However, the fictitious character of the limiting actions in these proposals may be an additional limitation.

This paper is focused on methods based on representative actions. An important advantage of these methods with respect to the ones based on limiting actions is that the DM may feel more comfortable setting representative actions than limiting profiles. Some of the former methods use a single "central" (or representative) action to characterize each class (e.g., Nemery and Lamboray, 2008; Almeida-Dias et al., 2010; Ishizaka and Nemery, 2014). Other methods use information about a few characteristics or representative actions (e.g., Köksalan et al., 2009; Almeida-Dias et al., 2012). Several approaches can handle many reference actions, which are not necessarily representative of their class (e.g., Jacquet-Lagrèze, 1995; Doumpos and Zopounidis, 2004; Fernandez and Navarro, 2011). In particular, we are interested in methods that use a single or a few characteristic actions to describe the related classes. Within this kind of approaches, ELECTRE TRI-nC (Almeida-Dias et al., 2012) deserves special recognition. ELECTRE TRI-nC is composed by two rules (ascending and descending procedures) that correspond via the transposition operation and are used conjointly. In this method, the information coming from "*x* outranks a representative action *r*" has the same value as the obtained from "a representative action *r* outranks *x*", and the overall information is used for assigning *x*. Both the ascending and descending rules fulfil a set of structural properties which are similar to those fulfilled by ELECTRE TRI-B and ELECTRE TRI-nB.

ELECTRE TRI-nC has evolved in two fundamental directions:

- A hierarchical ELECTRE TRI-nC with interacting criteria was proposed by Corrente *et al*. (2016). Ordinal classification problems can be solved in different levels of the hierarchy. The fulfilment of the structural properties (maybe altered by interaction of criteria) was not addressed by this paper.
- Recently, ELECTRE TRI-nC was extended to the interval framework by Fernández *et al*. (2020). The INTERCLASS-nC method exhibits symmetry with respect to the transposition operation and satisfies the same structural properties as the former method. This is interesting, since the outranking credibility index is calculated by INTERCLASS-nC in a quite different form[1]. So, we question whether the fulfilment of these properties depends on more general features.

In this paper, our aim is to study some axiomatic bases related to ordinal classification methods with similar features as ELECTRE TRI-nC: i) describing classes through their representative actions, and ii) exploiting an "at least as good" relation between actions to be assigned and representative actions.

Suppose any relation "*x* is at least as good as *y* with respect to a certain desirable property $\Xi$"; there is no matter the way to create this relation. Suppose that the pair DM-decision analyst wants to use such a relation to design a caracteristic actions-based assignment method to classes ordered in the sense of increasing $\Xi$. In this paper we address the following fundamental issue: To propose a general form of the decision rule, and identifying the requirements on the set of characteristic actions in order to fulfill: i) the entire set of structural properties, which should be similar to the ones suggested by Almeida-Dias *et al*. (2010); and ii) symmetry with respect to the transposition operation. This general characterization is inspired by ELECTRE TRI-nC, but it is more general.

The novelty of this paper is also supported by:

- To be an extension of ELECTRE TRI-nC and its desirable properties to any preference model under very general demands;
- To present a method to design new ordinal classification approaches with similar characteristics as ELECTRE TRI-nC, but based on diverse preference models.[2] On this base, the pair DM-decision analyst can choose the most appropriate preference model, then assessing the representative actions and using an ordinal classification method with the desirable features analyzed in this paper. This opens a very wide space for combining preference models and characteristic or representative actions in new ordinal classification methods.

The paper structure is the following: after this introduction, the method and its requirements are presented in Section 2. The fulfillment of the basic properties is proved in Section 3. Section 4 outlines different models of preference, which are compatible with the general method proposed here, and could bring new ordinal classification approaches. With two examples, the application of the method is illustrated in Section 5. Lastly, we give brief conclusions in Section 6.

---

[1] In "A hierarchical interval outranking approach with interacting criteria" (under review in European Journal of Operational Research), a hierarchical INTERCLASS-nC also satisfies the same set of structural properties.
[2] In "A theoretical look at ordinal classification methods based on comparing actions with limiting boundaries between adjacent classes" (under review in Annals of Operations Research), similar goals were reached concerning ELECTRE TRI-B and ordinal classification methods based on boundary actions.

**2 A generalized ordinal classification method based on representative actions**

**Definition 1. (Compatibility with the order of classes)**

Consider a set of $M$ ordered and predefined classes $C = \{C_1,\ldots,C_k,\ldots,C_M\}$, $(M \geq 2)$ (ordered in the sense of increasing a certain desirable feature $\Xi$). We say that a binary relation $S$ is compatible with the order of $C$ iff $xSy$ gives an argument to believe that $x$ should be classified into a class equal or higher than $y$.

**Condition 1 (Requirements on the relational system of preferences)[3]**

Let us consider a set of classes $C$ according to Definition 1, and a pair of binary relations $(D, S)$ with the following characteristics:

- $D$ is a transitive relation;
- $S$ is a reflexive relation, compatible with the order of $C$;
- $\forall\ (x, y, z) \in A \times A \times A$:
  i.   $xDy \Rightarrow xSy$;
  ii.  $xSy$ and $yDz \Rightarrow xSz$;
  iii. $xDy$ and $ySz \Rightarrow xSz$

If $S$ is transitive and reflexive, it suffices to take $D \equiv S$.

To fulfill a set of desirable properties the following demands are needed.

**Condition 2 (Characterization of categories through a reference set $R$)**

Each element $C_k$ belonging to the set of categories $C$ is characterized by a subset $R_k$ of representative actions $r_{k,j}$, $j=1,\ldots$ card $(R_k)$, fulfilling:

a.  For $k = 1, \ldots, M\text{-}1$, for all action $w$ in $R_k$, there is at least one action $z$ in $R_{k+1}$ such that $zDw$;
b.  For $k = 1, \ldots, M\text{-}1$, for all action $w$ in $R_{k+1}$, there is at least one action $z$ in $R_k$ such that $wDz$;
c.  For all $(k, h)$ $(h > k)$ $(k = 1, \ldots, M\text{-}1)$, for each action $w$ in $R_h$, there is no action $z$ in $R_k$ such that $zSw$.

The set $R = \{\ r_{k,j}, j=1,\ldots$ card $(R_k), k = 1, \ldots, M\ \}$ is called the reference set.

It should be remarked that Condition 2 is less demanding than the requirements in ELECTRE TRI-nC, in which each pair $(w, z) \in R_{k+1} \times R_k$ such fulfill $wDz$, and other stronger separability conditions in order to satisfy the whole set of structural properties of the method.

**Definition 2. ($S$-relation between actions and subsets of representative actions)**

a) $xSR_k$ if and only if there is $z \in R_k$ such that $xSz$;
b) $R_kSx$ if and only if there is $w \in R_k$ such that $wSx$;

---

[3] Under very general conditions, "at least as good as" relations created by multi-criteria decision methods and the Pareto dominance relation fulfill this demand.

**Proposition 1. (Properties of the *S*-relation between actions and subsets of representative actions)**

Suppose that the set *R* fulfills Condition 2. For all *x* belonging to *A*:

i. $xSR_k \Rightarrow xSR_h$ for all $h < k$;

ii. $R_kSx \Rightarrow R_hSx$ for all $h > k$.

The proof follows from Conditions 1, 2.a, and 2.b., and Definitions 2.a and 2.b.

Similarly to ELECTRE TRI-nC and the primal-dual procedure in (Bouyssou and Marchant, 2015) and Bouyssou *et al.*, 2020), we propose a method with two rules given by Definitions 3 and 4.

**Definition 3. (Descending assignment rule)**

Set *S*, *D* and *R* fulfilling Conditions 1 and 2. For all *x*, set $xSR_0$.

i. For $k = M, \ldots, 0$, determine the first $R_k$ such that $xSR_k$;

ii. If $k = M$, assign *x* to $C_M$;

iii. If $k=0$, assign *x* to $C_1$;

iv. For $0 < k < M$, select $C_k$ and $C_{k+1}$ as possible categories to assign *x*.

**Remark 1.**

a) If in Definition 3 *k* takes a value within the interval $1 \leq k < M$, $xSR_k$ supports assigning *x* to a class not worse than $C_k$; *not* ($xSR_{k+1}$) is an argument against assigning *x* to a class higher than $C_{k+1}$; combining these arguments, $C_k$ and $C_{k+1}$ are compatible assignment decisions;

b) *x* may be assigned to $C_k \Rightarrow xSR_h$ for all $h < k$ (Definition 3 and Proposition 1.i);

c) $xSR_k$ ($k \geq 1$) $\Rightarrow$ *x* may be assigned to $C_h$ for some $h \geq k$ (Definition 3 and Proposition 1.i);

d) *x* may be assigned to $C_k \Rightarrow not(xSR_h)$ for all $h > k+1$ (Definition 3 and Proposition 1.i);

e) $not(xSR_k) \Rightarrow$ *x* may be assigned to $C_h$ for some $h \leq k$ (Definition 3 and Proposition 1.i).

**Definition 4 (Ascending assignment rule)**

Set *S*, *D* and *R* fulfilling Conditions 1 and 2. For all *x*, set $R_{M+1}Sx$.

i. For $k = 1, \ldots, M+1$ determine the first $R_k$ such that $R_kSx$;

ii. If $k = 1$, assign *x* to $C_1$;

iii. If $k = M+1$, assign *x* to $C_M$;

iv. For $1 < k < M+1$, select $C_k$ and $C_{k-1}$ as possible categories to assign *x*.

In Definition 4 reasons similar to Remark 1.a, (but using instead $R_kSx$ and $not(R_{k-1}Sx)$), justify selecting $C_k$ and $C_{k-1}$ as possible assignments.

The above rules are inspired by the ones in ELECTRE TRI-nC (cf. Almeida-Dias *et al.*, 2012), but with a significant difference. Let us explain this in detail. Consider, for instance, the descending assignment rule. If $0 < k < M$, ELECTRE

TRI-nC uses the higher value of the selection function to choose a single assignment between $C_k$ and $C_{k+1}$. The selection function $\rho(x, R_k)$ is calculated by using the credibility indices $\sigma$ of the outranking between $x$ and the actions in $R_k$, combined with the "*min*" operator. In the present proposal, the binary relation $S$ does not necessarily come from a credibility degree of a binary preference, and we prefer to keep its generality.

The selection function in ELECTRE TRI-nC can be interpreted as a credibility degree of an indifference relation between $x$ and $R_k$. $\rho(x, R_{k+1}) > \rho(x, R_k)$ is a good reason to select $C_{k+1}$ as assignment for $x$ if the value $\rho(x, R_{k+1})$ reaches certain threshold. But when $\rho(x, R_{k+1})$ is close to zero, to choose the range $C_k$ - $C_{k+1}$ as possible assignment is a better option. To illustrate this point, consider an application of ELECTRE TRI-C with the following data:

Number of criteria: 4 (increasing preference)

Number of classes: 3 ($C_3$ is the most preferred)

Weights: (0.25, 0.25, 0.25, 0.25)

Indifference thresholds: (0, 0, 0, 0)

Preference thresholds: (1, 1, 1, 1)

Credibility threshold $\lambda = 0.75$

The single criterion with veto power is the fourth one; $v_4 = 3$

The classes are characterized by the following sets:

$R_1 = \{(4, 4, 4, 4)\}$; $R_2 = \{(7, 7, 7, 7)\}$; $R_3 = \{(10, 10, 10, 10)\}$

Let $x$ be the action (1, 4, 4, 7); we have $\sigma(x, r_3) = 0$, $\sigma(x, r_2) = 0.25$, $\sigma(x, r_1) = 0.75$; hence, $k=1$ according to the descending rule. Additionally, $\sigma(r_2, x) = 1$ and $\sigma(r_1, x) = 0$. Then, $\rho(x, R_1) = min\{\sigma(x, r_1), \sigma(r_1, x)\} = 0$, and $\rho(x, R_2) = min\{\sigma(x, r_2), \sigma(r_2, x)\} = 0.25 \Rightarrow x$ is assigned to $C_2$ by the descending rule of ELECTRE TRI-C. It is not difficult to analyze that $x$ is assigned also to $C_2$ by the ELECTRE TRI-C ascending procedure. However, $x$ is outranked by (respectively, outranks) $r_1$ in three criteria, whereas the action is worse than $r_2$ in all the criteria, except the fourth one; hence, a hesitation between $C_1$ and $C_2$ seems more appropriate, as suggested by Definition 3.

Note that the ascending and descending procedures in Definitions 3 and 4 correspond via the transposition operation. As discussed in Introduction, both rules should be used conjointly. However, since each rule suggests two possible classes, the range between the less preferred and the most preferred class could be inappropriately wide. Therefore, we explain below a way to reduce the range of classes when the ascending and descending rules are combined.

Firstly, we define some other binary relations between actions and subsets of representative actions.

**Definition 5. (Indifference, preference and incomparability relations between actions and representative subsets)**

**Indifference:** $xIR_k \Leftrightarrow xSR_k$ and $R_kSx$;

**Preference**: $xPR_k \Leftrightarrow xSR_k$ and $not(R_kSx)$; $R_kPx \Leftrightarrow R_kSx$ and $not(xSR_k)$

**Incomparability**: $xIncR_k \Leftrightarrow not(xSR_k)$ and $not(R_kSx)$

Thus, $x$ is comparable with $R_k$ if and only if $xSR_k$ or $R_kSx$. In this case, we say that $C_k$ is well-described with respect to action $x$. In order to achieve comparability between $x$ and $R_k$, it suffices to include in $R_k$ an action $z$ such that $zSx$ or $xSz$.

**Lemma 1. (Relationship between the assignments suggested by both rules)**

Suppose that $x$ is comparable with all $R_k$, $k =1, \ldots, M$. $x$ is assigned by the descending procedure to a possible class which is not lower than the highest possible class suggested by the ascending procedure.

*Proof:*

Suppose that $x$ is assigned to $C_1$ by the descending rule. Then, from Definition 3, $x$ does not outrank $R_h$ for $h = 1, \ldots, M$. From the premise of comparability, $R_1Sx$. Hence, $x$ is assigned to $C_1$ by the ascending rule (Definition 4).

The case when $x$ is assigned to $C_M$ is trivial.

Suppose now that $x$ is assigned to the range $C_k$ - $C_{k+1}$ by the descending rule. From Definition 3, $xSR_k$ and $not(xSR_h)$ for $h > k$. From the premise of comparability, it follows that $R_hSx$ for $h > k$. If in the ascending procedure $h = k + 1$ was the first index for which $R_hSx$, $x$ would be assigned to the range $C_k$ - $C_{k+1}$ (Definition 4). If $h < k + 1$, $x$ would be assigned to $C_{h-1}$ – $C_h$. This completes the proof.

**Remark 2.**

$xIR_h$ for $k \leq h \leq k'$ brings positive arguments to classify $x$ into a class within the range $C_k$-$C_{k'}$. The case where $xPR_h$ is combined with $xIR_{h+1}$ and $R_{h+2}Px$ suggests the assignment to $C_{h+1}$ as more appropriate.

On the other hand, $xIncR_h$ means lack of positive and negative arguments to suggest the assignment to $C_h$; $xPR_h$ combined with $xIncR_{h+1}$ do not contradict assigning $x$ to $C_h$ or $C_{h+1}$ because the action could be (from a preference point of view) close to the boundary between both classes.

Suppose that an action $x$ is assigned to $C_M$ by the descending rule and to $C_1$ by the ascending one.

From Definitions 3 and 4 we have $xSR_M$ and $R_1Sx$; From Propositions 1.i and 1.ii, it follows that $xSR_h$ and $R_hSx$ for $h=1,\ldots M \Rightarrow xIR_h$ for $h=1,\ldots M$. This justifies that $x$ may be assigned to any class in the interval $C_1$- $C_M$.

If $x$ is assigned to $C_M$ by the ascending rule and to $C_1$ by the descending procedure, the analysis is similar but replacing $I$ by $Inc$. Since $x$ is incomparable with $R_k$ for $k=1,\ldots, M$, there is no argument to assign $x$ to any particular class. All the assignments are possible.

Suppose that $x$ is assigned to $C_{k'}$ or $C_{k'-1}$ by the ascending procedure and to $C_{k''}$ or $C_{k''+1}$ by the descending rule. Let us analyze different situations:

**First case**: $C_{k'-1} = C_{k''}$ (Coincidence of the assignments)

From Definition 3 and Proposition 1.i, $x$ assigned to $C_{k''}$ or $C_{k''+1}$ by the descending rule $\Rightarrow xSR_{k''}$, $not(xSR_h)$ for $h>k''$, and $xSR_h$ for $h<k''$ \hspace{2em} (A)

From Definition 4 and Proposition 1.ii, $x$ assigned to $C_{k'}$ or $C_{k'-1}$ by the ascending rule $\Rightarrow R_{k'}Sx$, $not(R_hSx)$ for $h<k'$ and $R_hSx$ for $h>k'$ \hspace{2em} (B)

Combining A, B, and Definition 5, we have $xPR_h$ for $h \leq k''$ and $R_hPx$ for $h>k''$; $x$ may be assigned to $C_{k''}$ or $C_{k''+1}$ being (from a preference point of view) close to the border between these classes.

**Second case**: $k'=k''=K$ ($x$ is assigned to $C_{K-1}$ or $C_K$ by the ascending rule and to $C_K$ or $C_{K+1}$ by the descending one; the ranges of assignment intersect on $C_K$)

From Definition 3 and Proposition 1.i, $x$ assigned to $C_K$ or $C_{K+1}$ by the descending rule $\Rightarrow xSR_K$, $not(xSR_h)$ for $h>K$, and $xSR_h$ for $h<K$ \hspace{2em} (C)

From Definition 4 and Proposition 1.ii, $x$ assigned to $C_K$ or $C_{K-1}$ by the ascending rule $\Rightarrow R_K Sx$, $not(R_h Sx)$ for $h<K$ and $R_h Sx$ for $h>K$ \hfill (D)

Combining C, D, and Definition 5, it follows that $xIR_K$, $xPR_h$ ($h<K$) and $R_h Px$ for $h>K$. According to Remark 2, $x$ should be classified into $C_K$.

**Third case**: $k'<k''$ (the ranges of assignments do not intersect)

Combining A, B, and Definition 5, we have $xIR_h$ for $k'\leq h\leq k''$, $xPR_h$ for $h<k'$ and $R_h Px$ for $h>k''$. According to Remark 2, this result justifies the assignment to any class within the range $C_{k'} - C_{k''}$.

**Fourth case**: $k'-1=k''+1$ ($x$ is assigned to $C_{k'-1}$ or $C_{k'}$ by the ascending rule and to $C_{k''}$ or $C_{k''+1}$ by the descending one; the ranges of assignments intersect on $C_{k''+1} = C_{k'-1}$)

Combining A, B, and Definition 5, we obtain $xIncR_{k'-1}$, $xPR_h$ for $h<k'-1$ and $R_h Px$ for $h\geq k'$. According to Remark 2, any class within the range $C_{k''} - C_{k'}$ is a possible assignment for $x$.

**Fifth case:** $k'-1>k''+1$ (the ranges of assignments do not intersect)

From A, B, and Definition 5, it follows that $xIncR_h$ for $k''<h<k'$, $R_h Px$ for $h\geq k'$, and $xPR_h$ for $h\leq k''$. Again, according to Remark 2, any class within the range $C_{k''} - C_{k'}$ is a possible assignment for $x$.

For simplicity, when an action is assigned to a single class $C_k$ we say that the action is classified into the range $C_k - C_k$.

As a consequence of the analysis above, we propose the following rule for the conjoint use of the ascending and descending procedures:

**Definition 6. (Conjoint assignment rule)**

a) If an action $x$ is assigned to $C_M$ (respectively $C_1$) by the descending (respectively ascending) rule, take the range $C_1 - C_M$ as possible assignments for $x$;

b) If $x$ is assigned to the range $C_h - C_{h+1}$ by both ascending and descending rules, take the same range of classes as possible assignments for $x$;

c) Suppose that $x$ is assigned to the range $C_h - C_{k'}$ (respectively, $C_{k''} - C_{h'}$) by the ascending (resp. descending) procedure. Then, if $k''\geq k'$, take the range $C_{k'} - C_{k''}$ as possible assignments for $x$; otherwise ($k''<k'$), take the range $C_{k''} - C_{k'}$ as possible assignments for $x$.

Given an assignment range $C_k - C_h$, the classes $C_k$ and $C_h$ will be called the limits of the range.

Note that the limits of the conjoint assignment range are the lower limit of the descending assignment range and the upper limit of the ascending assignment range.

## 3 Structural properties of the method

Unlike precedent works as (Almeida-Dias et al., 2010, 2012) and (Fernández et al., 2020) in which the structural properties are analyzed for the ascending and descending rules, separately, here we also address the analysis of the properties regarding the conjoint assignment rule given by Definition 6.

Propositions 2-3 and Remark 3 describe the conformity property of the proposal.

**Proposition 2.**

The descending assignment rule fulfills the following properties:

i) Each $w \in R_M$ is assigned to $C_M$;

ii) Each $w \in R_k$ ($k<M$) is assigned to a class within the range $C_k - C_{k+1}$.

The proof is immediate from Definition 3 and Conditions 1 and 2.c.

**Proposition 3.**

The ascending assignment rule fulfills the following properties:

a) Each $w \in R_1$ is assigned to $C_1$;

b) Each $w \in R_k$ ($k>1$) is assigned to a class within the range $C_{k-1} - C_k$.

The proof is immediate from Definition 4 and Conditions 1 and 2.c.

**Remark 3.**

From Propositions 2 and 3, and Definition 6, it is obvious that each $z$ belonging to $R_k$ $k= 1,...M$, is assigned to $C_k$ by the conjoint assignment rule. We conclude that Conformity is not fulfilled by the separated rules, but it is satisfied by the conjoint assignment procedure.

**Proposition 4. (Monotonicity of the ascending and descending rules)**

i) If $y$ is assigned to a single $C_k$ by the descending (respectively, ascending) procedure and $xDy$, then $x$ is assigned to a single $C_{k'}$ ($k' \geq k$) or to a range $C_{k'} - C_{k'+1}$ ($k' \geq k$);

ii) If $y$ is assigned to a range $C_k - C_{k+1}$ by the descending (respectively, ascending) procedure and $xDy$, then $x$ is assigned to a single $C_{k'}$ ($k' \geq k+1$) or to a range $C_{k'} - C_{k'+1}$ ($k' \geq k$).

(See the proof in Appendix 1).

**Definition 7. (Comparison of ranges of classes)**

Let $Range_1 = C_k - C_{k'}$ and $Range_2 = C_h - C_{h'}$ be two ranges of classes in $C$. We say that $Range_1$ is not worse than $Range_2$ (denoted $Range_1 \geq Range_2$) iff $k \geq h$ and $k' \geq h'$.

**Proposition 5. (Monotonicity of the conjoint assignment procedure)**

If $y$ is assigned to $Range_1$ by the conjoint assignment procedure and $xDy$, then $x$ is assigned to a range $Range_2 \geq Range_1$.

(See the proof in Appendix 1).

**Definition 8 (Merging and splitting operations)**

(a) Merging operation: two adjacent categories, $C_k$ and $C_{k+1}$, will be merged to become a new one, $C'_k$, characterized by a new subset of reference actions, $R'_k = R_k \cup R_{k+1}$. The new set of classes is

$C= \{C_1,...,C_{k-1},C'_k,C_{k+2},...,C_M\}$, which, (updating the subscripts), can be denoted as $\{C'_1,...,C'_{k-1},C'_k,C'_{k+1},...,C'_{M-1}\}$. The new set of reference actions is $R= \{R_1,...,R_{k-1},R'_k,R_{k+2},...,R_M\}$, which can be denoted as $R'= \{R'_1,...,R'_{k-1},R'_k,R'_{k+1},...,R'_{M-1}\}$.

The fulfillment of Condition 2 of the new reference set $R'$ is a direct consequence of the fulfillment of Condition 2 on the previous (before merging) reference set $R$.

(b) Splitting operation: the category $C_k$ is split into two new adjacent classes, $C'_k$ and $C''_k$, characterized by two new distinct subsets of reference actions, $R'_k$ and $R''_k$ respectively. The new set of classes is $C = \{C_1,...,C_{k-1},C'_k,C''_k,C_{k+1},...,C_M\}$, which, (updating the subscripts), will be denoted as $\{C'_1,...,C'_{k-1},C'_k,C'_{k+1},C'_{k+2},...,C'_{M+1}\}$. The new set of reference actions is $R = \{R_1,...,R_{k-1},R'_k,R''_k,R_{k+1},...,R_M\}$, which will be denoted as $R' = \{R'_1,...,R'_{k-1},R'_k,R'_{k+1},R'_{k+2},...,R'_{M+1}\}$. $R'$ should fulfill Condition 2.

**Proposition 6. (Stability of the ascending and descending rules with respect to merging and splitting operations)**

The ascending and descending rules fulfill the following statements:

1. After a merging operation:
   a. an action that was classified into a range of classes which does not contain the merged categories is assigned to the same range;
   b. an action that was classified into a range of classes whose lower (respectively upper) category was merged is assigned to a range composed of the same upper (respectively lower) category, and the new class;
   c. when the merging is carried out with both classes of the range to which an action was assigned, this is classified into a range of classes which contains the new category.

2. After a splitting operation:
   a. an action that was classified into a range of classes which does not contain the modified category, keeps its assignment to the same range;
   b. an action that was classified into a range which contains the split class is either assigned to the range composed of the new classes, or to a range composed of a new class and the other class in the range before splitting.

(See the proof in Appendix 1).

**Proposition 7. (Analysis of the stability of the conjoint assignment rule with respect to merging and splitting operations)**

1. After a merging operation:
   a. an action, that was classified into a range of classes which does not contain the merged categories, keeps its assignment to the same range;
   b. an action, that was classified into a range of adjacent classes whose lower and upper limits are merged, is assigned to the new class;
   c. an action, that was classified into a range of classes whose lower (respectively upper) limit was merged, is assigned to a range composed of the same upper (respectively lower) limit, and the new class.

d. When the merging is carried out with two classes that are within the range (neither in the lower nor in the upper of the range) to which an action was assigned, it is classified into the same range as before the merging.

2. *After a splitting operation:*
   a. When the split class is adjacent to one of the limits of the assignment range of an action *x*, the new assignment range is the same, or the old limit is replaced by one of the new classes.
   b. When one of the limits of the assignment range is split, the new limit is one of the new classes, or one of their adjacent classes.
   c. When the split class is neither a limit of the assignment range, nor one of their adjacent classes, the assignment range keeps its limit classes.

(See the proof in Appendix 1).

**Remark 4.**

There are some cases (see 2.a and 2.b above) where, after the splitting, the new range of classes can have different limits from the original limits of the conjoint assignment range. It should be noticed that in such cases the union of $R'_k$ and $R'_{k+1}$ may be different from the old $R_k$. It is not surprising that, with different preference information, the assignment of an action could be modified.

**4 Examples of ordinal classification methods that can be built by using this proposal**

For each relational system (*D,S*) satisfying Definition 1 and Condition 1, an ordinal classification method can be built by using Definitions 3 and 4. Such a method would be symmetric with respect to the transposition operation, and it would fulfill the paradigmatic properties stated by Roy and Bouyssou (1993), and adapted by Almeida-Dias *et al*. (2010) for ELECTRE TRI-C. This remark is valid whatever the decision model, even beyond multi-criteria decision models. We will distinguish some important particular cases:

1. Classical ELECTRE framework

   This case arises when *S* is the a crisp outranking relation obtained from the credibility index of the outranking $\sigma(x,y)$ used by the later ELECTRE methods (Roy, 1991). Let $\delta$ denote a real number within ]0.5, 1], considered as a credibility threshold to establish the crisp outranking relations $xSy \Leftrightarrow \sigma(x,y) \geq \delta$; *D* is the classical Pareto dominance relation. It is well-known that such a relational system fulfills Condition 1. Setting the reference set fulfilling Condition 2, the ascending and descending rules and the conjoint assignment method from Definitions 3, 4, and 6 can be used. It would be a variant of ELECTRE TRI-nC, which fulfills similar properties, uses a more relaxed reference set, and suggests perhaps more justified assignments, as discussed in Section 2.

2. ELECTRE framework with interacting criteria

   Figueira *et al*. (2009) introduced a way to extend the ELECTRE family of methods to handle interactions among criteria. With such extension, these methods can now handle strengthening, weakening and antagonism between criteria. The concordance index, $c(x, y)$, is used by the ELECTRE methods to model the strength of the arguments in favor of the statement "action $x$ is at least as good as action $y$". $c(x, y)$ will now increase when there is a pair of criteria with strengthening interaction both supporting the statement; it will decrease when there are pairs of criteria with weakening interaction if both criteria support the statement. Finally, the concordance index will decrease when there is antagonism between criteria where one criterion supports the statement and the other opposes to it. Let $A$ be the set of actions, $G$ the set of criteria, and $g_i(x)$ denote the assessment of action $x \in A$ on criterion $g_i \in G$ such that, without loss of generality, $g_i(x)$ increases with preference. Then, the extended concordance index $c(x, y)$ is a non-decreasing function of $g_i(x) - g_i(y)$ for all criterion $g_i$ (Figueira *et al.*, 2009). The discordance index $d(x,y)$ in the extended version of ELECTRE is calculated from an aggregation of the marginal discordance indices, as traditionally. $d(x,y)$ requires that, for all $g_i$, $d(x,y)$ is non-increasing with respect to $g_i(x) - g_i(y)$. Through $c(x, y)$ and $d(x,y)$, a credibility index of the statement is calculated as $\sigma(x,y) = c(x, y) \cdot (1 - d(x,y))$. Finally, the outranking relation $S$ is defined as $xSy \Leftrightarrow \sigma(x,y) \geq \beta$, for a sufficiently great majority threshold $\beta$.

If $D$ is the Pareto dominance, then the pair $(D,S)$ fulfills Condition 1 (see Appendix 2). This indicates that ELECTRE TRI-nC can handle interacting criteria preserving its structural properties.

3. Multi-criteria ordinal classification based on a majority rule

    This kind of approaches build an outranking relation using a majority rule. As in (Meyer and Olteanu, 2019), a concordance measure is built with criterion weights and criterion scores; a crisp outranking relation $S$ is defined by comparison with a certain threshold. This relation satisfies Definition 1; if $D$ is the Pareto dominance, the relational system $(D,S)$ fulfills Condition 1. With appropriate setting of representative actions according to Condition 2, the method proposed in this paper can be used. This method fulfills the properties from Propositions 2-5.

4. Hierarchical ELECTRE

    ELECTRE methods that handle hierarchical structures can build an outranking relation $S_h$ on any non-elementary criterion $g_h$ (see e.g., Corrente *et al.*, 2013, 2016, 2017). This relation is reflexive and compatible with a set of ordered classes, as required by Definition 1. Taking $D$ as the Pareto dominance for sub-criteria of $g_h$, then such a relational system of preferences fulfills Condition 1, under the same claims given above. Thus, if the characteristic actions are defined properly regarding Condition 2, the method based on Definitions 3, 4, and 6 can work on the highest hierarchical level, or on any non elementary sub-criterion. The method fulfills the consistency properties from Propositions 2-5 and Remark 3. The hierarchical ELECTRE TRI-C with interacting criteria proposed by Corrente *et al.* (2016) is similar to our method, and hence it satisfies the same properties.

5. Interval outranking approach

    The so-called interval outranking approach was first proposed by Fernández *et al.* (2019) and later improved in (Fernández *et al.*, 2020). This approach extends the outranking paradigm to handle imprecisions on weights and

veto thresholds by using interval numbers. Imprecision and uncertainty related to criterion performances can also be handled by using interval numbers but, since different types of imprecision and uncertainty can affect criterion performances, the interval outranking approach can also deal with these situations by using discriminating thresholds. Which type of modeling tool should be used depends on the context of each criterion. Let $\sigma(x,y,\lambda)$ denote the credibility index of the interval outranking relation with a majority threshold $\lambda$. The outranking relation can be defined as $xSy \Leftrightarrow \sigma(x,y,\lambda) \geq \beta$. The outranking relation defined this way is compatible with a set of ordered classes (Definition 1). An interval-based dominance was described in (Fernández et al., 2019); it was proved in that work that the interval-based outranking and dominance relations fulfil Condition 1. Even when the interval-based outranking is not reflexive, it will always fulfill $xSx$ if action $x$ is characterized by pseudo-criteria or by real numbers; thus, satisfying the Conformity Property when the criterion performances of the characteristic actions in Condition 2 are real numbers. So, the method based on Definitions 3, 4 and 6 can be implemented, and it satisfies Propositions 2-5 and Remark 3. Indeed, the INTERCLASS-nC by Fernández et al. (2020) is a variant of the method proposed here.

6. Hierarchical interval outranking with interacting criteria

It is possible to extend the interval-based outranking approach to deal with criteria that interact and that are structured as hierarchies[4]. Let $S_h$ be an outranking relation related to a non-elementary criterion $g_h$. $S_h$ is compatible with a set of ordered classes with respect to $g_h$ (Definition 1). Now let $D$ be defined similarly to the interval dominance on the subset of elementary criteria descending from $g_h$, the pair $(D, S_h)$ fulfills Condition 1. Defining appropriately the representative actions of classes according to Condition 2, the method based on Definitions 3, 4, and 6 suggest assignments to classes related to $g_h$.

7. PROMETHEE TRI-nC

PROMETHEE calculates a binary preference degree, $\Pi(x, y)$, through which a reflexive binary relation compatible with a set of ordered classes can be defined as $xSy \Leftrightarrow \Pi(x, y) - \Pi(y, x) \geq 0$. Then, using the Pareto dominance, $D$, it is easy to demonstrate that the system $(D, S)$ satisfies Condition 1. A conjoint assignment method based on Definitions 3, 4, and 6 can be used.

8. Interval-based value functions

The preferences of the DM can be modeled by an interval value function $U$. The simplest function of this type is an interval weighted sum function. In such function, both criterion weights and criterion scores are interval numbers. Therefore, a reflexive outranking relation can be defined as $xSy \Leftrightarrow Poss(U(x) \geq U(y)) \geq 0.5$, where $Poss$ is the possibility function from Equation 1.

$$Poss(E \geq D) = \begin{cases} 1 & \text{if } p_{ED} > 1, \\ p_{ED} & \text{if } 0 \leq p_{ED} \leq 1, \\ 0 & \text{if } p_{ED} < 0 \end{cases} \quad (1)$$

---

[4] "A hierarchical interval outranking approach with interacting criteria" (under review in European Journal of Operational Research)

$E = [e^-, e^+]$ and $D = [d^-, d^+]$ are interval numbers and $p_{ED} = \frac{e^+ - d^-}{(e^+ - e^-)+(d^+ - d^-)}$.

In (Fernández *et al.*, 2019, 2020), the above possibility function is interpreted as a degree of credibility of $E \geq D$. So, the order relation on interval numbers is defined as $E \geq D \Leftrightarrow Poss(E \geq D) > 0.5$.

$S$ is reflexive and compatible with a set of ordered classes. Let us define $D$ as $xDy \Leftrightarrow Poss(U(x) \geq U(y)) \geq \alpha > 0.5$. Since the order relation on interval numbers is transitive (cf. Fernández *et al.*, 2019), the relational system $(D,S)$ satisfies Condition 1.

9. Compensatory preferences combined with veto conditions

   Assume the existence of a model where an ordinal value function $U$ is combined with the consideration of veto situations. This means that, for all $(x, y)$, $xSy \Leftrightarrow U(x) \geq U(y)$ and no veto condition is fulfilled $\Leftrightarrow x$ is at least as good as $y$. $S$ is reflexive and compatible with a set of ordered classes (Definition 1). Without loss of generality, suppose that $U$ is monotonically increasing with criterion scores. Again taking $D$ as the Pareto dominance relation, the pair $(D,S)$ fulfills Condition 1.

10. Decision models in group contexts

    If the property $\Xi$ is a measure of agreement, then properly defining $(D,S)$ the method based on Definitions 3, 4, and 6, could be used to assign potential collective decisions to classes of acceptable agreement.

## 5 Illustrative examples

Let us discuss two examples; the first one is the toy case illustrated in Section 2; the second example comes from a more realistic multi-criteria problem.

5.1 Revisiting the toy example in Section 2

This example corresponds to the classical ELECTRE framework; the action $x = (1, 4, 4, 7)$ should be assigned to one the three classes, described by the single actions $r_1 = (4, 4, 4, 4)$; $r_2 = (7, 7, 7, 7)$; $r_3 = (10, 10, 10, 10)$. As we discussed in Section 2, $x$ is assigned to $C_2$ by ELECTRE TRI-C.

With the outranking parameters described in Section 2, we had:

$\sigma(x, r_3) = 0$, $\sigma(x, r_2) = 0.25$, $\sigma(x, r_1) = 0.75$, $\sigma(r_3, x) = 1$, $\sigma(r_2, x) = 1$, $\sigma(r_1, x) = 0$.

The $S$-relation is taken as $xSy \Leftrightarrow \sigma(x,y) \geq \lambda$, where $\lambda$ is set as 0.75. Note that the set of representative actions fulfils Condition 2.

The binary relations that follow from Definition 5 are provided by Table 1.

**Table 1. Relations between $x$ and representative actions**

|   | $R_1$ | $R_2$ | $R_3$ |
|---|-------|-------|-------|
| $x$ | $P$ | $P^{-1}$ | $P^{-1}$ |

Note: $xP^{-1}R_k \Leftrightarrow R_kPx$

Applying the descending rule (Definition 3), $k=1$ is the first $k$ value for which $xSR_k$; hence, $x$ is assigned to the range $C_1$-$C_2$ by the rule. Applying the ascending rule, $k=2$ is the first value fulfilling $R_kSx$; action $x$ is assigned to the same range $C_1$-$C_2$ by the ascending rule. According to Definition 6.b, $x$ is assigned to the range $C_1$-$C_2$ by the conjoint method. We can argue that, with the information in Table 1, it is not possible to select a single class.

5.2 Assigning impact of R&D projects

The overall impact of a set of Research and Development projects is determined by four criteria: economic impact, social impact, scientific impact, and improvement of research competence. Each criterion is evaluated in a qualitative scale with levels { Very Low, Low, Below Average, Average, Above Average, High, Very High}.The DM does not want to accept projects with an overall impact lower than "High"; (s)he is considering to assign projects to one class of the ordered set $C$= {Unsatisfactory, High, Very High, Outstanding}.

The DM agrees on an ELECTRE model without veto conditions. Given the nature of the scale, (s)he sets discriminating thresholds equal to zero. Under these conditions, the model corresponds to a majority sorting rule. The DM assesses the same importance to all the criteria, so $w_i = 0.25$ for $i=1,…4$.

Table 2 shows a set of representative projects for each class. Note that these characteristic actions fulfil Condition 2.

**Table 2. The reference set**

| Project Id | Economic impact | Social impact | Scientific impact | Improvement of research competence | Overall Impact |
|---|---|---|---|---|---|
| 1 | High | High | High | Very High | Outstanding |
|  |  |  |  |  |  |
| 2 | Below Average | High | High | High | Very High |
| 3 | High | High | Below Average | High | Very High |
| 4 | High | Below Average | High | High | Very High |
| 5 | High | High | High | Below Average | Very High |
|  |  |  |  |  |  |
|  |  |  |  |  |  |
| 6 | High | Below Average | Below Average | Below Average | High |
| 7 | Below Average | High | Below Average | Below Average | High |
| 8 | Below Average | Below Average | High | Below Average | High |

| Project Id | Economic impact | Social impact | Scientific impact | Improvement of research competence | Overall Impact |
|---|---|---|---|---|---|
| 9 | Below Average | Below Average | Below Average | High | High |
| 10 | Average | Average | Average | Low | High |
|  |  |  |  |  |  |
|  |  |  |  |  |  |
| 11 | Low | Low | Low | Average | Unsatisfactory |
| 12 | Average | Low | Low | Low | Unsatisfactory |
| 13 | Low | Average | Low | Low | Unsatisfactory |
| 14 | Low | Low | Average | Low | Unsatisfactory |
| 15 | High | Very Low | Very Low | Very Low | Unsatisfactory |
| 16 | Very Low | High | Very Low | Very Low | Unsatisfactory |
| 17 | Very Low | Very Low | High | Very Low | Unsatisfactory |
| 18 | Very Low | Very Low | Very Low | High | Unsatisfactory |

For illustration purposes, let us consider the assignment of the projects in Table 3.

**Table 3. Projects to be assigned**

| Project Id | Economic dimension | Social dimension | Scientific impact | Improvement of research competence |
|---|---|---|---|---|
| 19 | High | High | High | High |
| 20 | High | High | Low | High |
| 21 | Above Average | Below Average | Above Average | Below Average |
| 22 | Above Average | Below Average | Low | Low |
| 23 | Average | Very High | Very High | Below Average |

Table 4 shows the preference (P), indifference (I), and incomparability (Inc) relations between projects and subsets of representative projects, according to Definitions 2 and 5.

**Table 4. Relations between projects and representative subsets**

|  | $R_1$ | $R_2$ | $R_3$ | $R_4$ |
|---|---|---|---|---|
| $p_{19}$ | P | P | I | I |
| $p_{20}$ | P | P | I | $P^{-1}$ |
| $p_{21}$ | P | I | $P^{-1}$ | $P^{-1}$ |
| $p_{22}$ | I | $P^{-1}$ | $P^{-1}$ | $P^{-1}$ |
| $p_{23}$ | P | P | P | Inc |

Note: $p_j P^{-1} R_k \Leftrightarrow R_k P p_j$

The assignments of the projects following Definitions 3, 4, and 6 are given by Table 5.

**Table 5. Assignments of projects**

|  | Descending rule (Def. 3) | Ascending rule (Def. 4) | Conjoint (Def. 6) |
|---|---|---|---|
| $p_{19}$ | Outstanding | High or Very High | Very High or Outstanding |
| $p_{20}$ | Very High or Outstanding | High or Very High | Very High |
| $p_{21}$ | High or Very High | Unsatisfactory or High | High |
| $p_{22}$ | Unsatisfactory or High | Unsatisfactory | Unsatisfactory |
| $p_{23}$ | Very High or Outstanding | Outstanding | Very High or Outstanding |

It should be remarked that the conjoint assignments are clearly justified by the relations *P* and *I* in Table 4.

## 6 Some conclusions

This paper has perhaps presented the widest generalization of the idea behind the ELECTRE TRI-nC multi-criteria ordinal classification approach, whatever the model used to represent the decision maker's preferences. Our proposal is a general approach to design ordinal classification methods based on comparing actions to be assigned with representative actions of their respective class. The main requirement is a relational system (*D,S*), where *S* is reflexive relation that is compatible with the preferential order of classes, and *D* is a transitive relation that should be a subset of *S*. Additionally, the proposal requires a reference set composed of representative actions of their classes. Unlike ELECTRE TRI-nC and related methods, our approach does not require any valued closeness relation or selection function. The proposed general approach is composed of two complementary assignment rules, which correspond through the transposition operation and should be used conjointly. These rules are similar to the ascending and descending procedures in ELECTRE TRI-nC, but a bit more conservative; instead of using a selection function as ELECTRE TRI-nC, the rules in this paper suggest a range of adjacent classes. The rules are combined in a less conservative conjoint assignment method, which suggests a range of classes determined by the lower (respectively, upper) limit of the descending (resp., ascending) assignment range. Under slight conditions on the representative actions (less demanding than in ELECTRE TRI-nC), the rules and the conjoint assignment procedure fulfill the structural properties of Conformity, Stability, Monotonicity, Homogeneity, Independence and Uniqueness, which are analyzed for each rule and for the conjoint assignment procedure. ELECTRE TRI-nC, its interval extension INTERCLASS-nC, and the hierarchical ELECTRE TRI-nC and INTERCLASS-nC with interacting criteria, can be considered as particular cases of the general approach proposed here. Using this, a plethora of ordinal classification procedures with desirable properties may arise from each decision model with capacity to build relations *S* and *D* fulfilling the basic features mentioned before. The pair DM-decision analyst can choose the most appropriate preference model, assessing the representative actions, and using an ordinal classification method with the desirable features analyzed in this paper. This opens a very wide space for combining preference models and characteristic or representative actions in new ordinal classification methods. This point was illustrated by many different kinds of decision models (Section 4); as an avenue of future research, some of them could be subject of forthcoming papers.

# APPENDIX 1

## Proofs of Propositions 4-7

**Proposition 4.**

*Proof:*

Descending rule:

Suppose that $y$ is assigned to $C_M$; from Definition 3, $ySR_M$; $xDy$ and $ySR_M \Rightarrow xSR_M$ (Condition 1 and Definition 2) $\Rightarrow x$ is assigned to $C_M$ (Definition 3)

Suppose that $y$ is assigned to $C_1$; from Definition 3 we have $not(ySR_1)$. Since $xDy$ there are two possibilities: $not(xSR_1)$ or $xSR_h$ for a certain $h \geq 1$. From Remark 1.d, $not(xSR_1) \Rightarrow x$ is assigned to $C_1$; $xSR_h$ ($h=M$) $\Rightarrow x$ is assigned to $C_M$ (Definition 3); $xSR_h$ ($1 \leq h < M$) $\Rightarrow x$ is assigned to the range $C_h - C_{h+1}$. Choose $k'=h$.

Suppose now that $y$ is assigned to the range $C_k - C_{k+1}$. From Definition 3, we have $ySR_k$ ($k \geq 1$); $xDy$ and $ySR_k \Rightarrow xSR_h$ for a certain $h \geq k$ from Condition 1 and Definition 2. If $h=M$, $x$ is assigned to $C_M$. If $h<M$, then $x$ is assigned to the range $C_h - C_{h+1}$. Taking $k'=h$ completes the proof.

Ascending rule:

The proof comes from the equivalence through the transposition operation.

**Proposition 5.**

*Proof:*

If $y$ is assigned to the range $C_1$-$C_M$ the proof is immediate from Definition 6.a , Definition 7, and Proposition 4.

If $y$ is assigned to the range $C_h$-$C_{h+1}$ by both ascending and descending rules, the proof is immediate from Definition 6.b , Definition 7, and Proposition 4.

Consider now the case where $y$ is assigned to the range $C_{k'-1}$-$C_{k'}$ by the ascending procedure., and to $C_{k''}$-$C_{k''+1}$ by the descending rule. Let us analyze first the case $k'' \geq k'$. According to Definition 6.c, $y$ is assigned to the range $C_{k'}$- $C_{k''}$. From Proposition 4, $x$ is assigned to $C_{h'-1}$-$C_{h'}$ by the ascending rule ($h' \geq k'$), and to the range $C_{h''}$-$C_{h''+1}$ by the descending rule ($h'' \geq k''$).

If $h'' \geq h'$, from Definition 6.c, it follows that $x$ is assigned to $C_{h'}$- $C_{h''}$. Since $h' \geq k'$ and $h'' \geq k''$, we have $(C_{h'} - C_{h''}) \geq (C_{k'} - C_{k''})$.

If $h'' < h'$, from Definition 6.c $x$ is assigned to $C_{h''}$- $C_{h'}$. Since $h'' \geq k''$, it is obvious that $(C_{h''} - C_{h'}) \geq (C_{k'} - C_{k''})$.

Lastly, let us analyze the case $k''<k'$. From Definition 6.c, $y$ is assigned to the range $C_{k''}$- $C_{k'}$.

If $h''\geq h'$, $h'\geq k' \Rightarrow (C_{h''}-C_{h'}) \geq (C_{k''}-C_{k'})$;

If $h''<h'$, $(h''\geq k''$ and $h'\geq k') \Rightarrow (C_{h''}-C_{h'}) \geq (C_{k''}-C_{k'})$.

The proof is complete.

**Proposition 6.**

*Proof:*

Descending rule:

   1.a.

   Suppose that $x$ was assigned to the range $C_i$-$C_j$ ($j=i+1$ or $j=i$) before the merging,

   If $i>k+1$ or $j<k$, from Definition 3 and Condition 2.c, we have that $x$ is assigned to the same range of classes.

   1.b.

- Suppose that $x$ was assigned to the range $C_{k+1}$-$C_{k+2}$ before the merging From Definition 3 we have that $xSR_{k+1} \Rightarrow xS(R_k \cup R_{k+1}) \Rightarrow xSR'_k$; also, we know that $not(xSR_n)$ for all $n>k+1 \Rightarrow$ we have that $x$ is assigned to the range $C'_k$ - $C'_{k+1}$, where $C'_{k+1}$ is the old $C_{k+2}$.

- Suppose that $x$ was assigned to the range $C_M$-$C_M$; from Definition 3 we have that $xSR_M$, and $xSR_M$ with Definition 8.a $\Rightarrow xSR'_{M-1} \Rightarrow x$ is assigned to the range $C'_{M-1}$ - $C'_{M-1}$, where $C'_{M-1}$ is the old $C_M$.

- Suppose that $x$ was assigned to the range $C_{k-1}$-$C_k$ ; from Definition 3, it follows that $xSR_{k-1}$ and $not(xSR_n)$ for all $n\geq k \Rightarrow not(xSR'_k) \Rightarrow$ we have that $x$ is assigned to the range $C'_{k-1}$ - $C'_k$, where $C'_{k-1}$ is the old $C_{k-1}$.

- Suppose that $x$ was assigned to the range $C_1$-$C_1$ (where $k=1$ in Definition 8.a); from Definition 3 we have that $xSR_0$ and $not(xSR_n)$ for all $n>0 \Rightarrow not(xSR'_n)$ for all $n>0$. Hence, $x$ is assigned to the range $C'_1$ - $C'_1$.

   1.c.

- Suppose that $x$ was assigned to the range $C_k$-$C_{k+1}$ before the merging. From Definition 3, it follows that $xSR_k \Rightarrow xS(R_k \cup R_{k+1}) \Rightarrow xSR'_k$; also, from Definition 3, $not(xSR_n)$ for all $n>k \Rightarrow not(xSR'_n)$ for all $n>k \Rightarrow x$ is assigned to the range $C'_k$ - $C'_{k+1}$ (Definition 3), where $C'_{k+1}$ is the old $C_{k+2}$ (if $k+1=M \Rightarrow x$ is assigned to the range $C'_k$ - $C'_k \equiv C'_{M-1}$ - $C'_{M-1}$).

   2.a.

   Suppose that $x$ was assigned to the range $C_i$-$C_j$ ($j=i+1$ or $j=i$) before the splitting;

   If $i>k$ or $j<k$, from Definition 3 and Condition 2.c, we have that $x$ is assigned to the same range of categories.

   2.b.

- Suppose that $x$ was assigned to the range $C_k$-$C_{k+1}$ before the splitting. From Definition 3, it follows that $xSR_k$ and $not(xSR_{k+1}) \Rightarrow not(xSR'_{k+2})$ (Definition 8.b). From Proposition 1 i. we have that $xSR_{k-1} \Rightarrow xSR'_{k-1}$. There are the following three possible cases:

    1. $xSR'_k$ and $xSR'_{k+1} \Rightarrow$ (Definition 3) $x$ is assigned to the range $C'_{k+1}$-$C'_{k+2}$.

    2. $xSR'_k$ and $not(xSR'_{k+1}) \Rightarrow$ (Definition 3) $x$ is assigned to the range $C'_k$-$C'_{k+1}$.

    3. $not(xSR'_k)$ and $not(xSR'_{k+1}) \Rightarrow$ (Definition 3) $x$ is assigned to the range $C'_{k-1}$-$C'_k$.

- Suppose that $x$ is assigned to the range $C_{k-1}$-$C_k$; from Definition 3, it follows that $xSR_{k-1} \Rightarrow xSR'_{k-1}$ and $not(xSR_k)$; also, $not(xSR_{k+1}) \Rightarrow not(xSR'_{k+2})$ (Definition 8.b). Thus, we have the same three above possible cases with the same conclusions.

Note:
- If $k=M$, the range $C_k$-$C_{k+1}$ is $C_M$-$C_M$ and the range $C'_{k+1}$-$C'_{k+2}$ is $C'_{M+1}$-$C'_{M+1}$.
- If $k=1$, the range $C_{k-1}$-$C_k$ is $C_1$-$C_1$.

Ascending rule:

The proof comes from the equivalence through the transposition operation.

The proof is complete.

**Proposition 7.**

*Proof:*

1.a

Suppose that $x$ was assigned to the range $C_h$-$C_l$ before the merging.

- Suppose that $h > k+1 \Rightarrow$ from Definition 8.a and Proposition 1 we have both $not(R_mSx)$ and $xSR_m$ for all $m < h$.
- Suppose that $l < k \Rightarrow$ from Definition 8.a and Proposition 1 we have both $R_mSx$ and $not(xSR_m)$ for all $m > l$.

From the above analysis, we conclude that the new category does not intervene in the classification; therefore, the action is classified to the same range after the merging.

1.b

- Suppose that $x$ was assigned to the range $C_{k-1}$-$C_k$ by the ascending rule and $C_{k+1}$-$C_{k+2}$ by the descending rule.
  $\Rightarrow R_kSx$ and $not(R_mSx)$ for all $m < k$ (Definition 4) $\Rightarrow R'_k Sx$ (Definition 8(a) and Definition 2.b) and $not(R'_mSx)$ for all $m < k \Rightarrow x$ is assigned to the range $C'_{k-1} - C'_k$ by the ascending rule.
  $\Rightarrow xSR_{k+1}$ and $not(xSR_m)$ for all $m > k+1$ (Definition 3) $\Rightarrow xSR'_k$ (Definitions 8.a and 2.a) and $not(xSR'_m)$ for all $m > k \Rightarrow x$ is assigned to the range $C'_k - C'_{k+1}$ by the descending rule.
  $\Rightarrow x$ is assigned to the range $C'_k$ by the conjoint assignment rule after the merging.
- Suppose that $x$ was assigned to the range $C_k$-$C_{k+1}$ by the ascending rule and $C_k$-$C_{k+1}$ by the descending rule (the same range).
  $\Rightarrow x$ is assigned to the range $C'_k$ by the conjoint assignment rule after the merging.

1.c

Suppose that $x$ was assigned to the range $C_{k+1}$-$C_h$ (or $C_k$-$C_h$) before the merging.

- Suppose that $x$ was assigned to the range $C_k$-$C_{k+1}$ ($C_{k-1}$-$C_k$) by the ascending rule and $C_h$-$C_{h+1}$ by the descending rule.
  $\Rightarrow R_{k+1}Sx$ ($R_kSx$) and $not(R_mSx)$ for all $m < k+1$ ($m < k$) (Definition 4) $\Rightarrow R'_k Sx$ (Definition 8(a) and Definition 2.b) and $not(R'_mSx)$ for all $m < k \Rightarrow x$ is assigned to the range $C'_{k-1} - C'_k$ by the ascending rule.
  $\Rightarrow xSR_h$ and $not(xSR_m)$ for all $m > h$ (Definition 3) $\Rightarrow x$ is assigned to the same range $C_h - C_{h+1}$ ($C'_{h-1} - C'_h$) by the descending rule.
  $\Rightarrow x$ is assigned to the range $C'_k$-$C'_{h-1}$ by the conjoint assignment rule after the merging.
- Suppose that $x$ was assigned to the range $C_{h-1}$-$C_h$ by the ascending rule and $C_{k+1}$-$C_{k+2}$ ($C_k$-$C_{k+1}$) by the descending rule.

⇒ $R_hSx$ and $not(R_mSx)$ for all $m<h$ (Definition 4) ⇒ $x$ is assigned to the same range $C_{h-1}$– $C_h$ ($C'_{h-2}$ – $C'_{h-1}$) by the ascending rule.

⇒ $xSR_{k+1}$ ($xSR_k$) and $not(xSR_m)$ for all $m>k+1$ ($m>k$) (Definition 3) ⇒ $xSR'_k$ (Definition 8(a) and Definition 2.a) and $not(xSR'_m)$ for all $m>k$ ⇒ $x$ is assigned to the range $C'_k$ – $C'_{k+1}$ by the descending rule.

⇒ $x$ is assigned to the range $C'_k$-$C'_{h-1}$ by the conjoint assignment rule after the merging.

Suppose that $x$ was assigned to the range $C_h$-$C_k$ before the merging.

The proof is similar to the previous case.

1.d

Suppose that $x$ was assigned to the range $C_h$-$C_l$ before the merging where $h<k$ and $l>k+1$.

- Suppose that $x$ was assigned to the range $C_{h-1}$-$C_h$ by the ascending rule and $C_l$-$C_{l+1}$ by the descending rule.
  ⇒ $R_hSx$ and $not(R_mSx)$ for all $m<h$ (Definition 4) ⇒ the assignment of $x$ by the ascending rule is independent of the merging.
  ⇒ $xSR_l$ and $not(xSR_m)$ for all $m>l$ (Definition 3) ⇒ the assignment of $x$ by the descending rule is independent of the merging.
  ⇒ $x$ is assigned to the same range after the merging.
- Suppose that $x$ was assigned to the range $C_{l-1}$-$C_l$ by the ascending rule and $C_h$-$C_{h+1}$ by the descending rule.
  ⇒ $R_lSx$ and $not(R_mSx)$ for all $m<l$ (Definition 4) ⇒ $not(R_{k+1}Sx)$ and $not(R_kSx)$ ⇒ $not((R_k \cup R_{k+1})Sx)$ ⇒ the assignment of $x$ by the ascending rule is independent of the merging.
  ⇒ $xSR_h$ and $not(xSR_m)$ for all $m>h$ (Definition 3) ⇒ $not(xSR_k)$ and $not(xSR_{k+1})$ ⇒ $not(xS(R_k \cup R_{k+1}))$ ⇒ the assignment of $x$ by the descending rule is independent of the merging.
  ⇒ $x$ is assigned to the same range after the merging.

2.a

We know that the limits of the conjoint assignment range are the lower limit of the descending assignment range and the upper limit of the ascending assignment range. This feature will be used in the proof.

Suppose that $C_h$ is the (upper and/or lower) limit of the range to which $x$ is assigned, and $C_k$ (the class that is split) is adjacent to $C_h$.

- If $k=h+1$ and $x$ was assigned to the range $C_h$-$C_{h+1}$ by the descending rule ⇒ $xSR_h$ y $not(xSR_m)$ for all $m>h$ (Definition 3).
  - If $xSR'_{k+1}$ ⇒ the new limit is $C'_{k+1}$.
  - If $xSR'_k$ and $not(xSR'_{k+1})$ ⇒ the new limit is $C'_k$.
  - If $not(xSR'_k)$ and $not(xSR'_{k+1})$ ⇒ the new limit is $C'_h=C_h$.
  
  In the case in which $x$ was assigned to the range $C_{h-1}$-$C_h$ by the ascending rule ⇒ $R_hSx$ and $not(R_mSx)$ for all $m<h$ (Definition 4) ⇒ $R'_kSx$ and $R'_{k+1}Sx$ (Definition 8(b) and Proposition 1.ii) ⇒ the new limit is $C'_h=C_h$.

- If $k=h-1$ and $x$ was assigned to the range $C_h$-$C_{h+1}$ by the descending rule ⇒ $xSR_h$ and $not(xSR_m)$ for all $m>h$ (Definition 3) ⇒ $xSR'_k$ and $xSR'_{k+1}$ (Definition 8(b) and Proposition 1.i) ⇒ the new limit is $C'_{h+1}=C_h$.

  In the case that $x$ was assigned to the range $C_{h-1}$-$C_h$ by the ascending rule ⇒ $R_hSx$ and $not(R_mSx)$ for all $m<h$ (Definition 4).
  - If $R'_kSx$ ⇒ the new limit is $C'_k$.

- Si *not(R'$_k$Sx)* and *R'$_{k+1}$Sx* ⇒ the new limit is *C'$_{k+1}$*.
- Si *not(R'$_k$Sx)* and *not(R'$_{k+1}$Sx)* ⇒ the new limit is *C'$_{h+1}$=C$_h$*.

2.b

Suppose that $C_k$ is the split (upper and/or lower) limit of the range to which *x* was assigned.

- If *x* was assigned to the range *C$_k$-C$_{k+1}$* by the descending rule ⇒ *xSR$_k$* and *not(xSR$_m$)* for all *m>k* (Definition 3).
  - If *xSR'$_{k+1}$* ⇒ the new limit is *C'$_{k+1}$*
  - If *xSR'$_k$* and *not(xSR'$_{k+1}$)* ⇒ the new limit is *C'$_k$*
  - If *not(xSR'$_k$)* and *not(xSR'$_{k+1}$)* the new limit is *C'$_{k-1}$=C$_{k-1}$*
- If *x* was assigned to the range *C$_{k-1}$-C$_k$* by the ascending rule ⇒ *R$_k$Sx* and *not(R$_m$Sx)* for all *m<k* (Definition 4) and *R$_n$Sx* for all *n>k* (Proposition 1(b)).
  - If *R'$_k$Sx* ⇒ the new range is *C'$_{k-1}$-C'$_k$* ⇒ the new limit is *C'$_k$*
  - If *not(R'$_k$Sx)* and *R'$_{k+1}$Sx* ⇒ the new range is *C'$_k$-C'$_{k+1}$* ⇒ the new limit is *C'$_{k+1}$*
  - If *not(R'$_k$Sx)* and *not(R'$_{k+1}$Sx)* ⇒ the new range is *C'$_{k+1}$-C'$_{k+2}$* ⇒ the new limit is *C'$_{k+2}$=C$_{k+1}$*.

2.c

Suppose that $C_h$ is the limit (upper and/or lower) of the range to which *x* was assigned $C_k$ is the split class.

- If *k>h+1*
  - If *x* was assigned to the range *C$_h$-C$_{h+1}$* by the descending rule ⇒ *xSR$_h$* and *not(xSR$_m$)* for all *m>h* (Definition 3) ⇒ *not(xSR$_{h+1}$)* ⇒ *not(xSR'$_k$)* and *not(xSR'$_{k+1}$)* (Definition 8(b)) ⇒ the new limit is *C'$_h$=C$_h$*.
  - If *x* was assigned to the range *C$_{h-1}$-C$_h$* by the ascending rule ⇒ *R$_h$Sx* and *not(R$_m$Sx)* for all *m<h* (Definition 4) ⇒ *R'$_k$Sx* and *R'$_{k+1}$Sx* (Definition 8(b)) ⇒ the new limit is *C'$_h$=C$_h$*.
- If *k<h-1*
  - If *x* was assigned to the range *C$_h$-C$_{h+1}$* by the descending rule ⇒ *xSR$_h$* and *not(xSR$_m$)* for all *m>h* (Definition 3) ⇒ *xSR$_{h-1}$* ⇒ *xSR'$_k$* and *xSR'$_{k+1}$* (Definition 8(b)) ⇒ the new limit is *C'$_{h+1}$=C$_h$*.
  - If *x* was assigned to the range *C$_{h-1}$-C$_h$* by the ascending rule ⇒ *R$_h$Sx* and *not(R$_m$Sx)* for all *m<h* (Definition 4) ⇒ *not(R$_{h-1}$Sx)* ⇒ *not(R'$_k$Sx)* and *not(R'$_{k+1}$Sx)* (Definition 8(b)) ⇒ the new limit is *C'$_{h+1}$=C$_h$*.

The proof is complete.

## APPENDIX 2

### Proof of Condition 1 in the ELECTRE framework with interacting criteria

*i. xDy ⇒ xSy*

*xDy* ⇒ *g$_i$(x)≥g$_i$(y)* ∀*g$_i$∈G* ⇒ *c(x,y)=1* and *d(x,y)=0* ⇒ *σ$_I$(x,y)=1* ⇒ *xSy*

*ii. xSy* and *yDz* ⇒ *xSz*

*yDz* ⇒ *g$_i$(y)≥g$_i$(z)* ∀*g$_i$∈G* ⇒ *-g$_i$(y)≤-g$_i$(z)* ∀*g$_i$∈G* ⇒ *g$_i$(x)-g$_i$(y)≤g$_i$(x)-g$_i$(z)* ∀*g$_i$∈G*. Since *c(x,y)* is non-decreasing with respect to *g$_i$(x)-g$_i$(y)*, then

$c(x,z) \geq c(x,y)$ ............................................................................................................. (a)

Since $g_i(x)-g_i(y) \leq g_i(x)-g_i(z) \ \forall g_i \in G$ and $d(x,y)$ is non-increasing with respect to $g_i(x)-g_i(y)$, then

$d(x,z) \leq d(x,y)$ ............................................................................................................. (b)

From (a), (b) and the definition of $\sigma_I(x, y)$ we have that $\sigma_I(x, y) \leq \sigma_I(x, z) \Rightarrow xSz$.

*iii. xDy and ySz* $\Rightarrow xSz$

$xDy \Rightarrow g_i(x) \geq g_i(y) \ \forall g_i \in G \Rightarrow g_i(x)-g_i(z) \geq g_i(y)-g_i(z) \ \forall g_i \in G$. Since $c(x,y)$ is non-decreasing with respect to $g_i(x)-g_i(y)$ we have

$c(x,z) \geq c(y,z)$ ............................................................................................................. (c)

Since $g_i(x)-g_i(z) \geq g_i(y)-g_i(z) \ \forall g_i \in G$ and $d(x,y)$ is non-increasing with respect to $g_i(x)-g_i(y)$, then

$d(x,z) \leq d(y,z)$ ............................................................................................................. (d)

From (c), (d) and the definition of $\sigma_I(x, y)$, it follows that $\sigma_I(y, z) \leq \sigma_I(x, z) \Rightarrow xSz$.